# Sculpturing sound fields with real-space topology

Qing Tong[1] and Shubo Wang[1,*]

[1]Department of Physics, City University of Hong Kong, Tat Chee Avenue, Kowloon, Hong Kong, China.

*Corresponding author: Shubo Wang (shubwang@cityu.edu.hk)

**ABSTRACT**

Artificial structures have been widely used to manipulate sound fields. Most properties of these structures derive from the material and geometry. Few are explicitly related to the structural topology in the real space. Here, we discover a fundamental connection between the real-space topology of acoustic structures and the topological properties of sound fields. We find that the genus of acoustic cavities can protect the birth of singularities in velocity polarization and isopressure line fields. The total topological index of the surface singularities is equal to the cavities' Euler characteristic. This intriguing relationship is rooted in the Poincaré–Hopf theorem and is irrelevant to the specific material, geometric details, or excitation properties. The isopressure line singularities lead to acoustic hotspots and quiet zones. The velocity polarization singularities give rise to nontrivial polarization Möbius strips and skyrmion textures. The results enable sound sculpturing with structural topology and can find applications in acoustic communications, acoustic sensing, and noise control.

## INTRODUCTION

The past two decades have witnessed the booming research on artificial structures for wave manipulation, such as metamaterials and metasurfaces(*1–4*). These structures are usually realized by designing the specific geometry of the building block units. They can carry unusual properties not available in nature, which give rise to many intriguing phenomena, including negative refraction(*5, 6*), cloaking(*7, 8*), helical dichroism(*9–11*), and flat lens imaging(*12–14*). Recently, artificial structures have been used as a convenient platform to explore the topological properties of classical waves such as light and sound(*15–18*). This has developed into two main streams: the topology associated with the momentum space and the topology associated with the real space. The former has uncovered numerous intriguing phenomena, such as one-way edge states(*15, 19–22*), non-Abelian quaternion charges(*23, 24*), and topological lasing(*25*). The latter has enabled the realization of unconventional structured fields, including knots(*26–29*), skyrmions(*30, 31*), and toroids(*32, 33*). The two streams are both concerned with the topology of wave fields, while the real-space topology of structures and its effects on wave properties remain rarely explored.

Although airborne sound is a longitudinal wave, it can carry vector properties through the velocity field. Akin to electromagnetic fields, the velocity field of a generic sound wave in three-dimensional space is elliptically polarized and can be expressed as $\mathbf{v}(\mathbf{r}) = [\mathbf{A}(\mathbf{r}) + i\mathbf{B}(\mathbf{r})]e^{i\theta(\mathbf{r})}$, where $\mathbf{A}$ and $\mathbf{B}$ are the major and minor axes of the polarization ellipse. The distribution of the polarization ellipses in real space can give rise to interesting configurations(*34, 35*). Specially, they can form topological defects known as polarization singularities(*34–37*). Of particular interest are the C points (where the field is circularly polarized and the major axis $\mathbf{A}$ of the polarization ellipse is ill-defined) and V points (where the field amplitude vanishes and the field direction is undetermined)(*38*). The C points carry nonzero acoustic spin density $\mathbf{S} \propto \mathrm{Im}(\mathbf{v}^* \times \mathbf{v})$(*39–41*) that can generate intriguing chiral sound-matter interactions(*42–44*). The V points are higher order polarization singularities formed of degenerate C points(*45*). Both are topological singularities of the line field $\mathbf{A}$ (since $\mathbf{A}$ and $-\mathbf{A}$ correspond to the same polarization axis). Topological singularities also exist in the scalar pressure field $p(\mathbf{r})$. Since the isopressure lines can be considered a line field, its spatial distribution can give rise to singularities at which the orientation of the line field is ill-defined. These singularities correspond to the local maxima (i.e., hot spots) or minima (i.e., quiet zones) of pressure.

Here, we show that the topological properties of the two types of singularities (i.e., the singularities of the velocity polarization field and isopressure line field) in acoustic cavities are fundamentally dictated by the real-space topology of the cavities as a result of the Poincaré–Hopf (PH) theorem and are irrelevant to the specific material or geometric details or excitation properties. These singularities carry rich topological properties that can enable sound sculpturing with the real-space topology of acoustic structures.

**RESULTS**

**Topological sound fields in the cavities with different real-space topology**

We consider airborne sound waves inside a spherical cavity of radius $R$ with a hard boundary, as shown by the inset in Fig. 1(A). The cavity is excited by a plane wave port at a small circular opening on the $+x$ axis, as indicated by the blue arrow in Fig. 1(A). The sound speed inside the cavity is set to be $v = v_0(1 + 0.05i)$ with $v_0 = 343$ m/s. We conduct full-wave simulation of the system by using a finite-element package COMSOL and determine the polarization singularities of the complex velocity field $\mathbf{v}$ at the frequency $f = 150$ Hz for $R = 1.05$ m. The results are shown in Fig. 1(A), where the orange/green lines denote C lines and red/blue dots denote V points. As seen, there are two ring-shape C lines emerging inside the cavity. Each C line can be characterized by a topological polarization index defined as $I_{\text{pl}} = 1/(4\pi) \oint d\varphi$, where $\varphi$ is the azimuthal angle of the polarization state on the Poincaré sphere, and the integral is evaluated on a small loop enclosing the C line. The orange (green) C line has $I_{\text{pl}} = +1/2$ ($I_{\text{pl}} = -1/2$). The polarization index can be intuitively understood by examining the polarization ellipses near the C lines, as shown in Fig. 1(D) and 1(E) for the polarization ellipses on the $xoy$-plane, where the line segments inside the ellipses denote the major axis $\mathbf{A}$. The background color in Fig. 1(D) and 1(E) shows the phase of the scalar field $\mathbbm{v} = \mathbf{v} \cdot \mathbf{v}$. Clearly, the C points correspond to the phase singularities of $\mathbbm{v}$. Accordingly, we can also define a topological phase index for each C line: $I_{\text{ph}} = 1/(2\pi) \oint \nabla \text{Arg}[\mathbbm{v}] \cdot d\mathbf{r}$. This phase index is directly related to the polarization index as $I_{\text{pl}} = \text{sign}(\mathbf{t} \cdot \mathbf{S}) I_{\text{ph}}/2$ with $\mathbf{t}$ being a tangent unit vector of the C line and $\mathbf{S}$ being the local spin density(45). In Fig. 1(A), there are also two V points (denoted by the red dots) emerging on the cavity's surface, both carrying the index $I_{pl} = +1$. Figure 1(F) shows the polarization major axes **A** (denoted by the black arrows) field near the V point marked as "F" in Fig. 1(A). The background color denotes the phase $\text{Arg}[\mathbb{v}]$. In contrast to the C points, the V points do not correspond to the phase singularity of $\mathbb{v}$.

Is the emergence of the polarization singularities accidental? What decides the indices of these polarization singularities? It turns out that the answers are rooted in the subtle relationship between the topological properties of velocity field and the real-space topology of sphere. Under the hard boundary condition, the **A** field near the spherical surface is everywhere tangent to the surface. Thus, it can be considered as a line field defined on a two-dimensional smooth manifold (i.e., the spherical surface). According to the PH theorem, this tangent line field must have discrete singularities, and the total index of the singularities must be equal to the Euler characteristic of the manifold $\chi = 2 - 2g$, where $g$ is the genus (i.e., number of "holes") of the manifold(*46*). In the spherical cavity in Fig. 1(A), there are only two V points on the surface, and the sum of their indices is $\sum I_{pl} = +1 \times 2 = +2$, indeed agreeing with the Euler characteristic of sphere $\chi = 2 - 0 = 2$. For the C lines inside the cavity, they are not in touch with the spherical surface, and thus are irrelevant to the topology of the cavity.

To verify the above interpretation, we simulated the polarization singularities emerging in the acoustic cavities with different Euler characteristics $\chi$ under the same excitation. The results are shown in Fig. 1(B) for a torus cavity with genus $g = 1$ and in Fig. 1(C) for a double-torus cavity with genus $g = 2$. In the torus case, there are four V points (denoted by the red and blue dots) born on the torus surface, among which two have the index $I_{pl} = +1$ and the other two have the index $I_{pl} = -1$. Figure 1(G) shows the polarization major axes **A** near the negative V point labeled by "(G)" in Fig. 1(B). In addition, there are six C lines emerging inside the torus cavity, among which four connect to the surface and two form isolated rings inside the cavity without touching the surface. Thus, there are a total of 12 polarization singularities on the torus surface with a total index $\sum I_{pl} = +1 \times 2 - 1 \times 2 + \frac{1}{2} \times 4 - \frac{1}{2} \times 4 = 0$, which agrees with the Euler characteristic of the torus $\chi = 2 - 2 \times 1 = 0$. For the double torus in Fig. 1(C), there are eight V points emerging on the surface, among which two have the positive index $I_{pl} = +1$ and six have the negative index $I_{pl} = -1$. In addition, there are six C lines with $I_{pl} = +1/2$ and four C lines with $I_{pl} = -1/2$ inside the cavity, and they all connect to the surface. Therefore, on the surface of the double torus, there are a total of 28 polarization singularities with a total index $\sum I_{\text{pl}} = +1 \times 2 - 1 \times 6 + \frac{1}{2} \times 12 - \frac{1}{2} \times 8 = -2$, which is equal to the Euler characteristic of the double torus $\chi = 2 - 2 \times 2 = -2$. These results demonstrate that the emergence of the surface polarization singularities is protected by the structural topology of the acoustic cavity. The global topological properties (i.e., total polarization index) of these singularities are solely decided by the Euler characteristic of the cavity, and thus only depends on the number of "holes" in the cavity's surface. This fundamental relationship between the total index of velocity-field polarization singularities and the real-space topology of acoustic cavities remains valid for closed cavities and lossless cavities (see Supplementary Note 1).

Apart from the velocity field, the real-space topology also decides the topological properties of the pressure field. Figure 2(A-C) shows the isopressure contours (denoted by the white lines) and the pressure amplitude (denoted by the background color) near the surface of the cavities in Fig. 1. Since the isopressure lines can be considered a line field defined on the cavity surface, they also have topological singularities emerging on the surface as a result of the PH theorem, as denoted by the black dots. Figure 2(D-F) show the isopressure line patterns near the three singularities marked in Fig. 2(A-C), which carry the index $I = +1, -1,$ and $-2$, respectively. On the spherical surface, there are only two singularities with positive index $I = +1$. There are 24 singularities on the torus surface and 60 singularities in the double-torus surface. Importantly, we find that the total index of these surface singularities are $\sum I = 1 \times 2 = 2$ for the sphere, $\sum I = 1 \times 13 + (-1) \times 9 + (-2) \times 2 = 0$ for torus, and $\sum I = 1 \times 25 + (-1) \times 23 + (-2) \times 2 = -2$ for double-torus. Therefore, they agree with the Euler characteristics of the cavities and satisfy the PH theorem. In addition, we note that the singularities of isotropressure line field correspond to the local maxima and minima of the pressure. Figure 2(G) shows the phase of pressure near the two minima marked by black arrows in Fig. 2(C), which corresponds to two phase singularities. In other words, the real-space topology of the acoustic cavities guarantees the existence of acoustic "hot spots" and "quite zones" near the cavity surface. In a real-life scenario, for a room constructed with smooth walls (corresponding to the hard boundaries) and a door (corresponding to the small opening of the cavity), one can always find locations near the room walls where the outdoor sound is strongly suppressed (i.e., quite zones) or enhanced (i.e., hot spots). This can give rise to interesting applications in outdoor noise control and indoor-outdoor communications. Naturally, there may also exist pressure maxima/minima inside the cavities induced by interference, such as the blue strip regions in Fig. 2(A-C) corresponding to the nodes of the standing waves, but they are not protected by the real-space topology and can disappear at variations of cavity geometry or excitations. In the following, we will focus on the velocity field in the cavities due to its rich vector and topological properties.

**Effects of symmetry and geometric singularity**

Symmetry plays an essential role in deciding the nontrivial properties of conventional topological systems. To explore the effects of geometric symmetry on the properties of velocity polarization singularities, we consider the cavities in Fig. 3(A) and 3(B) with genus $g = 0$, which are excited by a plane wave port located at the opening marked by the blue arrow. The cavity in Fig. 3(A) has mirror symmetry with respect to the *xoy*- and *xoz*-plane, while the cavity in Fig. 3(B) has no symmetry. As seen, the total number of polarization singularities and their configurations are different from that of the spherical cavity in Fig. 1(A). There are four C lines connecting to the surface in both cases, among which two C lines have the index $I_{pl} = +1/2$, and the other two C lines have the index $I_{pl} = -1/2$. Additionally, there are two V points of $I_{pl} = +1$ emerging on the surface. Consequently, the total index of the singularities on the cavity surfaces is identical to that of the spherical cavity: $\sum I_{pl} = +2$, satisfying the PH theorem. These results show that, while the symmetry of the acoustic cavities can affect the number and distribution of the polarization singularities, the global topological properties (i.e., total index) of the surface polarization singularities are invariant against continuous deformations of the cavity geometry and are irrelevant to the symmetry of the cavity.

One necessary condition for the application of the PH theorem in the considered systems is that the cavities' surfaces must be smooth. What will happen if geometric singularities are introduced into the surfaces? To address this question, we consider the cavity shown in Fig. 3(C) with genus $g = 0$, which has a singular boundary introduced by removing a cylindrical portion from the original spherical cavity in Fig. 1(A). Under the same excitation by the plane wave port, there are three C lines connecting to the surface and one V point at the opening, and their index sum is $\sum I_{pl} = \frac{1}{2} \times 2 - \frac{1}{2} \times 4 + 1 = 0$. In addition, there are two "quasi-V points" appearing at the locations marked by the red circles in Fig. 3(C). Figures 3(E) and 3(F) show the polarization major axes **A** near the two quasi-V points on the two surfaces jointed at the singular boundary. The polarization indices of the quasi-V points are ill-defined due to the absence of a local tangent plane. Thus, the total index of the polarization singularities is not necessarily equal to the Euler characteristic of the cavity. On the other hand, if the singular boundary is smoothed out, as shown in Fig. 3(D), the two quasi-V points turn into two V points with the positive index $I_{pl} = +1$. Figures 3(G) and 3(H) show the polarization major axes **A** near the two V points. Clearly, this smooth boundary case still satisfies the PH theorem.

**Evolutions of polarization singularities**

While the global topological property (i.e., total index) of the polarization singularities is independent of the excitation frequency, their detailed configurations can change when the frequency varies, giving rise to topological evolutions such as merging and bifurcation. To study these evolutions, we simulate the polarization singularities emerging in the spherical cavity and torus cavity when the frequency increases from 125 Hz to 230 Hz and from 140 to 160 Hz, respectively. Figure 4(A-E) shows the singularities and the phase $\text{Arg}[\mathbb{v}]$ in the spherical cavity. We notice that, as the frequency increases, two C lines of opposite polarization indices (corresponding to the orange and green circles) are born inside the cavity and gradually separate from each other. The orange C line with $I_{pl} = +1/2$ expands but does not touch the surface of the cavity. The green C line with $I_{pl} = -1/2$ shrinks and merges into a V point (marked by the blue dot in Fig. 4(E)). Interestingly, this V point carries an anisotropic polarization index: $I_{pl} = -1$ on any mirror plane (e.g., *xoy*-plane) and $I_{pl} = +1$ on the *yz*-plane (see Supplementary Note 2). Importantly, during the evolution from Fig. 4(A) to 4(E), the total index of the surface polarization singularities is invariant since the C rings are born in pairs and do not touch the cavity surface.

Figure 4(F-J) shows the polarization singularities and the phase $\text{Arg}[\mathbb{v}]$ in the torus cavity when the frequency increases from 140 Hz to 160 Hz. Initially, there are two pairs of C lines with opposite indices $I_{pl} = \pm 1/2$ emerging inside the cavity and two pairs of V points with indices $I_{pl} = \pm 1$ emerging on the cavity surface. As the frequency increases, we notice that the ends of the two C lines with $I_{pl} = -1/2$ (marked by $\xi_1$ and $\xi_2$) reach the inner surface of the torus, and each C line converges to a V point of index $I_{pl} = -1$ on the torus surface. Meanwhile, there are two C rings (marked by $\xi_3$ and $\xi_4$) newly born inside the torus, and each C ring is formed of a pair of C lines with opposite polarization index ($I_{pl} = \pm 1/2$) joined at two ends. Each pair of the C lines further splits and connects to the torus surface. Obviously, the total polarization index of the surface singularities is conserved and agrees with the Euler characteristic of the torus $\chi = 0$, even though the total number of the surface singularities varies during this evolution. These evolutions further demonstrate the robust relationship between the structural topology and the topological properties of the velocity field.

**Polarization Möbius strips**

The C lines protected by the real-space topology can give rise to intriguing polarization Möbius strips inside the cavities. We consider the major axes **A** evolving on closed loops in the yellow plane as depicted in Fig. 5(A), which forms an angle $\theta$ with the *xoy*-plane. We find that, for any nonzero value of $\theta$, the major axes **A** varies along the loops and can form Möbius strips. Figures 5(D) and 5(E) show the major axes **A** (denoted by the arrows) evolving along two loops encircling two different C lines, i.e., the orange ($I_{pl} = +1/2$) and green ($I_{pl} = -1/2$) C lines in Fig. 5(A), respectively. The color of the arrows denotes the phase Arg[$\mathbb{v}$]. The large yellow arrows on the C lines denote the direction of the C lines defined as $\mathbf{t}_c = \text{sign}(I_{ph})\mathbf{t}$, where **t** is a tangent unit vector of the C lines. In both cases, the Möbius strips manifest a nontrivial topology with a half-turn twist. The number of twists and the twisting direction are decided by the phase index $I_{ph}$ of the C lines. The Möbius strips in Fig. 5(D) and 5(E) twist a half turn in opposite directions because the phase indices of the C lines surrounded by the strips are $I_{ph} = +1$ and $I_{ph} = -1$, respectively (assuming the positive direction is along the +z direction). To further understand the relationship between the Möbius strips and the C lines, we consider the polarization major axes **A** evolving on the loops enclosing two C lines, as schematically shown in Fig. 5(B), where the orange (green) dots denote the C points with polarization index $I_{pl} = +1/2$ ($I_{pl} = -1/2$). Figure 5(F) shows the polarization configuration on the solid-line loop in Fig. 5(B), where the polarization strip is topologically trivial with no twist. Figure 5(G) shows the polarization configuration on the dashed-line loop in Fig. 5(B), which is also topologically trivial with no twist. In both cases, the total phase index of the enclosed two C lines is zero, while the total polarization index is +1 for the solid-line loop and 0 for the dashed-line loop. These results demonstrate that the topological properties of velocity-field polarization on closed loops are directly determined by the phase index $I_{\text{ph}}$. To further verify this point, we consider the polarization configurations in the double-torus cavity in Fig. 1(C). We plot the major axes **A** on the chosen dashed-line (solid-line) loop in Fig. 5(C), which encloses two C points with the same (opposite) polarization index, and the results are shown in Fig. 5(H) [Fig. 5(I)]. Interestingly, in both cases, we notice a second-order Möbius strip (i.e., the strip twists two half-turns along the loop). This is because that the enclosed two C lines in Fig. 5(H) (and 5(I)) have the same direction $\mathbf{t}_c$ (i.e., pointing downward as denoted by the yellow arrows) and carry the same phase index $I_{\text{ph}} = -1$ (assuming the positive direction is $+z$) as shown by the phase $\text{Arg}[\mathbb{v}]$, resulting in a total phase index of -2 for each chosen loop in Fig. 5(C). Thus, we conclude that the topology (number of twists) of the polarization Möbius strips of the velocity field is decided by the phase indices of all the C lines enclosed by the loop.

**Skyrmion and skyrmionium textures**

In addition to the Möbius strips, the spatial variation of the velocity vector **v** associated with the polarization singularities can give rise to interesting topological textures, including skyrmions and skyrmionium. The skyrmions are swirling vortex-like configurations of vector field originally proposed in particle theory(*47*) and subsequently observed in condensed matter systems such as chiral magnetic materials(*48*, *49*). Recently, skyrmion textures have been realized in optical and acoustic fields through the interference of multiple surface waves(*30*, *31*). The skyrmioniums exhibit donut-like textures composed of two skyrmions with opposite topological numbers(*50*, *51*) and have not been explored in acoustic systems. Interestingly, both skyrmion and skyrmioniums can appear in the velocity fields associated with the polarization singularities in our considered systems. Figure 6 (a-c) show the normalized velocity fields on *yz*-planes inside the spherical cavity, which clearly correspond to the Néel-type skyrmion and skyrmioniums(*30*, *31*, *52*, *53*). They can be characterized by the skyrmion number $Q = \frac{1}{4\pi} \iint \mathbf{n} \cdot (\partial_y \mathbf{n} \times \partial_z \mathbf{n}) \, dydz$ (*31*, *35*, *54*), where $\mathbf{n} = \text{Re}(\mathbf{v} \cdot \mathbf{e}^{-i\omega t})/|\text{Re}(\mathbf{v} \cdot \mathbf{e}^{-i\omega t})|$ denotes the normalized velocity vector and the integral is evaluated on the *yz*-plane on which the skyrmion/skyrmionium locates. For the skyrmions in Fig. 6(A) and 6(B), we obtain $Q = -1$ and $Q = +1$, respectively. The right insets show the rotation of the normalized velocity vectors **n** along the radial direction. For the skyrmionium in Fig. 6(C) we obtain $Q = 0$. We note that the skyrmions and skyrmioniums are time-dependent, and their skyrmion number $Q$ can change with time (see Supplementary Note 3). In addition, such topological textures of the velocity field can also emerge in lossless cavities (see Supplementary Note 3).

The emergence of the skyrmion and skyrmionium textures of the velocity field is closely related to the C lines and V points inside the cavity. Figure 6(D) and 6(E) show the phase $\text{Arg}[\mathbb{v}]$ on the *xoy*-plane, which exhibits discontinuity (jumping from $-\pi$ to $+\pi$) due to the presence of C lines. The white dashed lines mark the positions of the skyrmions in Fig. 6(A) and 6(B). The arrows denote the directions of the local normalized velocity vectors, which are linearly polarized on the spherical surface and on the center axis (i.e., *x* axis) due to the hard boundary condition and the mirror symmetry of the sphere. The direction of the normalized velocity vector evolves continuously inside the cavity except at the V points and at the phase discontinuity lines on the surface and center axis, where it changes sign abruptly. This explains the opposite directions of the normalized velocity vector at the skyrmions' center and boundary. Figure 6(F) shows the phase $\text{Arg}[\mathbb{v}]$ on the *xoy*-plane corresponding to the case of Fig. 6(C), where the dashed line marks the position of the skyrmionium. The emergence of the skyrmionium can be attributed to the spatial evolutions of the velocity vector associated with the appearances of more phase discontinuity lines near boundaries induced by more C lines emerging at the high frequencies.

**DISCUSSION**

In summary, we uncover an intrinsic relationship between the topological properties of sound fields and the structural topology of acoustic cavities. We find that the total topological indices of the velocity polarization singularities and the isopressure line singularities born on the surface are solely determined by the Euler characteristic of the cavities and are invariant against the continuous deformations of the cavities. The underlying mechanism is attributed to the nature of the sound fields near the structure's surface—both the velocity polarization and the isopressure contours are line fields defined on smooth manifolds and thus are governed by the PH theorem. This enables sound sculpturing with the real-space topology of structures, leading to intriguing spatial configurations of the velocity and pressure fields inside the cavities. The pressure field exhibits topologically protected hot spots and quiet zones, which can be applied to achieve efficient indoor-outdoor communications and noise suppression. The velocity field can carry nonzero spin density due to the C points, providing a robust mechanism for generating chiral sound-matter interactions. The velocity field can also give rise to polarization Möbius strips of various orders as well as skyrmion and skyrmionium textures, offering rich degrees of freedom for encoding information and manipulating small particles. Our study expands the realm of topological acoustics from the momentum space to the real space, opening exciting possibilities of exploring the real-space topological properties of acoustic fields.

**MATERIALS AND METHODS**

**Numerical simulations**

Full-wave numerical simulations are performed by using the finite-element package COMSOL Multiphysics. We assumed a hard boundary condition at the cavities' surfaces. All the cavities are excited by a plane wave port at the small opening (marked by the blue arrow in the figures) with the incident direction along -$x$ direction. The speed of sound is set to be $v = v_0(1 + 0.05i)$ for the lossy cavities with $v_0 = 343$ m/s. We consider three cavities in Fig. 1. The spherical cavity has a radius of $R = 1.05$m and a small circular opening with a radius of $a = 0.1$ m. The torus cavity has a major radius of $R_1 = 1$ m, a minor radius of $R_2 = 0.55$ m, and a small elliptical opening with a major radius of $r_1 = 0.18$m and a minor radius of $r_2 = 0.1$m. The double-torus cavity has a major radius of $R_1 = 1$ m and a minor radius of $R_2 = 0.55$ m and a small opening (same scales as in the torus cavity). The distance between the two holes is $d = 2$ m.

To explore the impact of symmetry on the acoustic polarization singularities, we study singularities in the cavities in Fig. 3(A) and 3(B). The ellipsoid in Fig. 3(A) has three axes of 1.25 m, 0.85 m, and 0.85 m, as well as a small opening with a major radius of 0.18 m and a minor radius of 0.12 m. The irregular cavity in Fig. 3(B) has a radius of 0.8 m and a small opening with a radius of 0.1 m. Regarding the influence of geometric singularities, we investigate the polarization singularities emerging in the incomplete spherical cavities (with a cylindrical part removed) with and without smooth edges in Fig. 3(C) and 3(D). These two incomplete spherical cavities both have radii of 1.05m, and the removed cylinders both have radii of 0.5 m. Both cavities have an opening with a radius of 0.1 m.

**Acknowledgments**: The work described in this paper was supported by grants from the Research Grants Council of the Hong Kong Special Administrative Region, China (Projects No. CityU 11306019 and No. AoE/P-502/20) and National Natural Science Foundation of China (No. 12322416).

**Author contributions**: S.W. conceived and supervised the project. Q.T. performed the numerical simulations. All authors contributed to the analysis and discussion of the results and manuscript preparation.

**Competing interests**: The authors declare no competing interests.

**Data and materials availability**: The authors declare that all data supporting the findings of this study are available within the paper and its Supplementary Information files. Additional data related to this paper are available from the corresponding authors upon reasonable request.

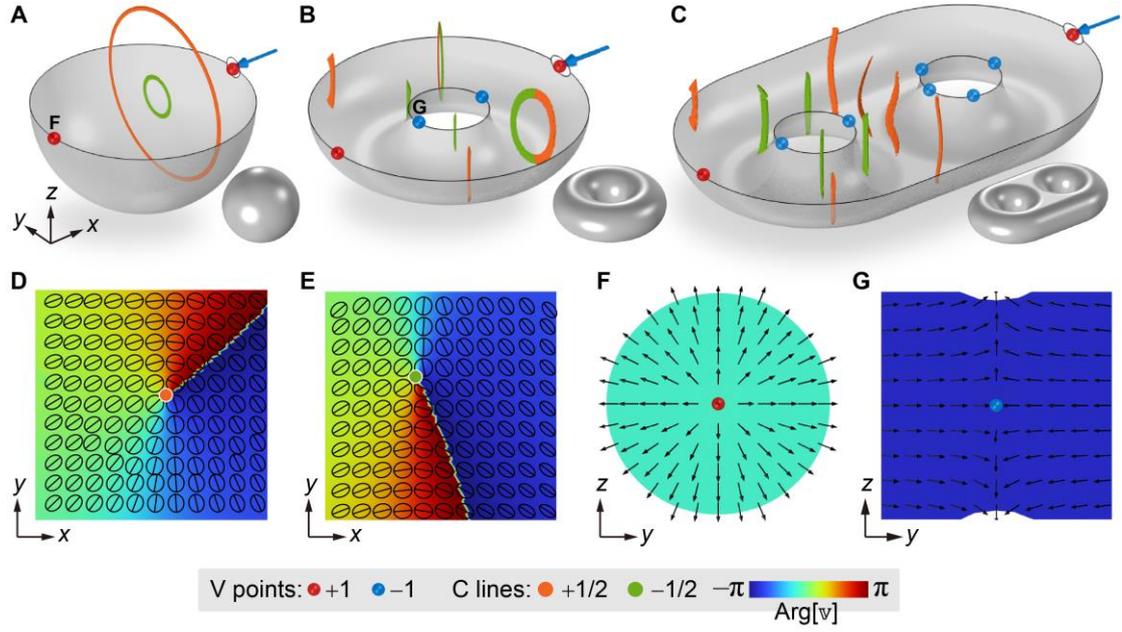

**Figure 1. Velocity polarization singularities emerging in acoustic cavities with different genus.** The C lines and V points emerging in (**A**) spherical cavity with genus $g = 0$, (**B**) torus cavity with genus $g = 1$, and (**C**) double-torus cavity with genus $g = 2$. A half of the cavities' surfaces is removed to show the internal distribution of the polarization singularities. The blue arrows indicate the direction of the incident acoustic plane wave with frequency $f = 150$ Hz. The polarization ellipses and major axes **A** near the C points with index (**D**) $I_{pl} = +1/2$ and (**E**) $I_{pl} = -1/2$ in the sphere system. The polarization major axes **A** near the V points with index (**F**) $I_{pl} = +1$ and (**G**) $I_{pl} = -1$, corresponding to the points marked as "F" and "G" in the sphere and torus systems. The background color in (**D-G**) denotes the phase Arg $[\mathbb{v}]$ = Arg $[\mathbf{v} \cdot \mathbf{v}]$.

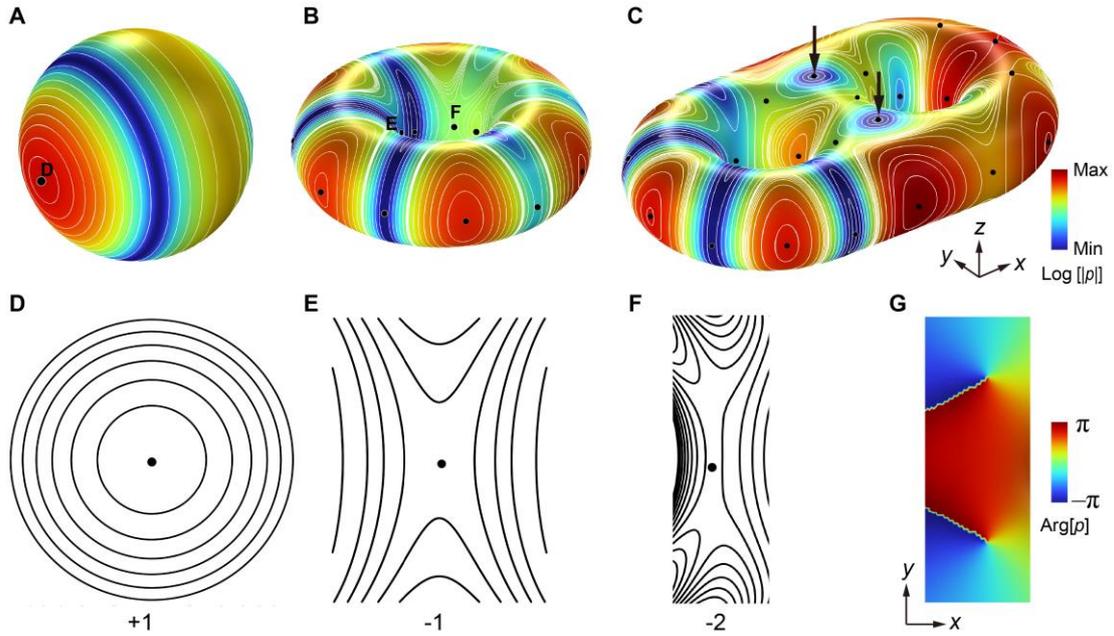

**Figure 2. Singularities of isopressure line field emerging in acoustic cavities with different genus.** Pressure and isopressure lines on the surface of (**A**) the spherical cavity, (**B**) the torus cavity, and (**C**) the double-torus cavity. The local isopressure line fields near the marked singularities are shown in (**D**) with index +1, (**E**) with index -1, and (**F**) with index -2. (**G**) The phase of the pressure field near the two singularities marked by two black arrows in (**C**), showing two phase singularities coinciding with the singularities of the isopressure line field.

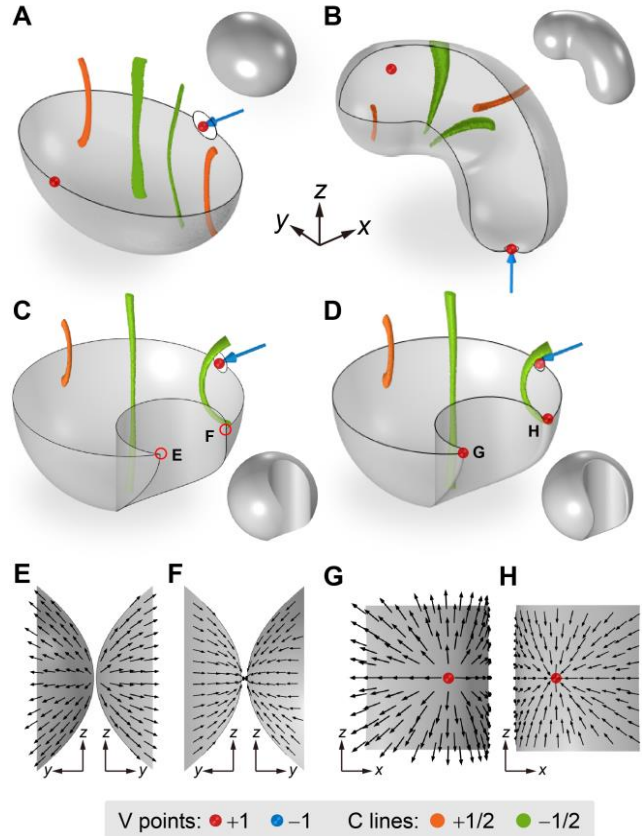

**Figure 3. Effects of geometric symmetry and geometric singularity.** (**A**) Polarization singularities emerging in an ellipsoid cavity with mirror symmetries. (**B**) Polarization singularities emerging in an irregular cavity without symmetry. (**C**) Polarization singularities emerging in a cavity with a singular boundary. (**D**) Polarization singularities emerging in a cavity with the singular boundary smoothed out. (**E**), (**F**) Polarization major axes **A** near the two quasi-V points marked in (**C**). (**G**), (**H**) Polarization major axes **A** near the two V points marked in (**D**).

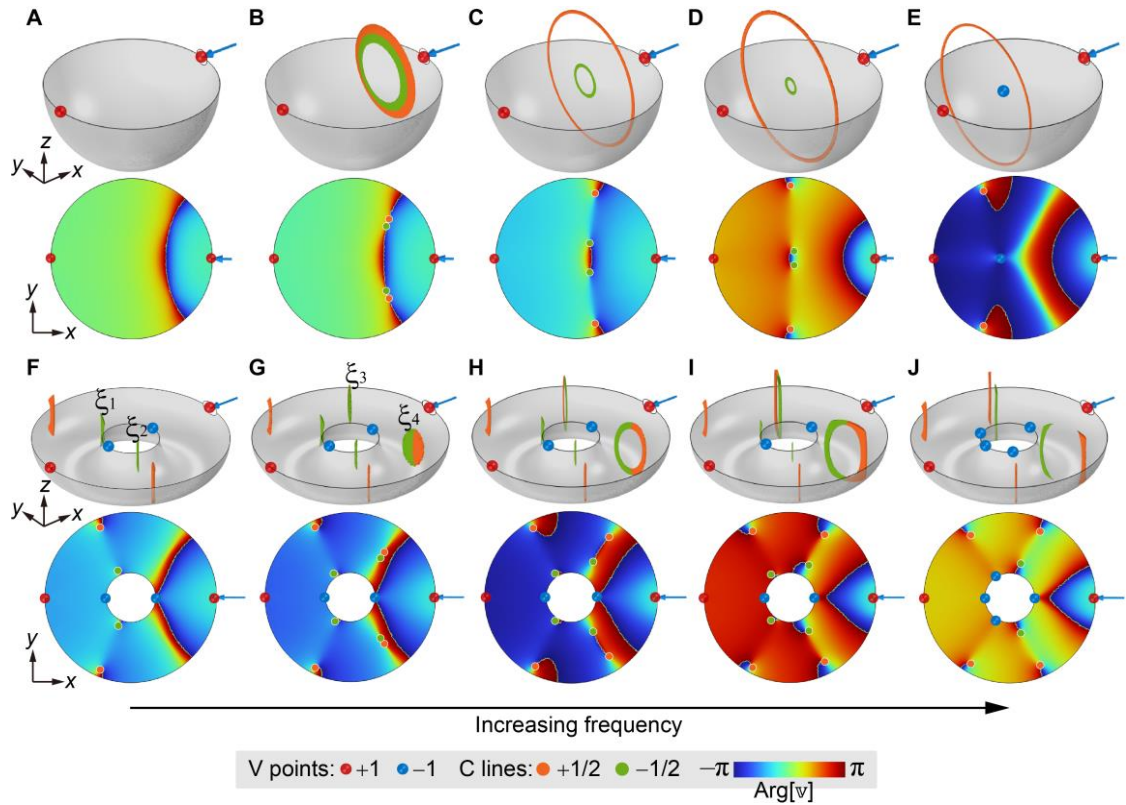

**Figure 4. Evolutions of polarization singularities**. The C lines and V points as well as the phase Arg[$\mathbb{v}$] at different frequencies: (**A**) 125 Hz, (**B**) 130 Hz, (**C**) 150 Hz, (**D**) 180 Hz, (**E**) 230 Hz for the spherical cavity, and (**F**) 140 Hz, (**G**) 145 Hz, (**H**) 150 Hz, (**I**) 155 Hz, (**J**) 160 Hz for the torus cavity.

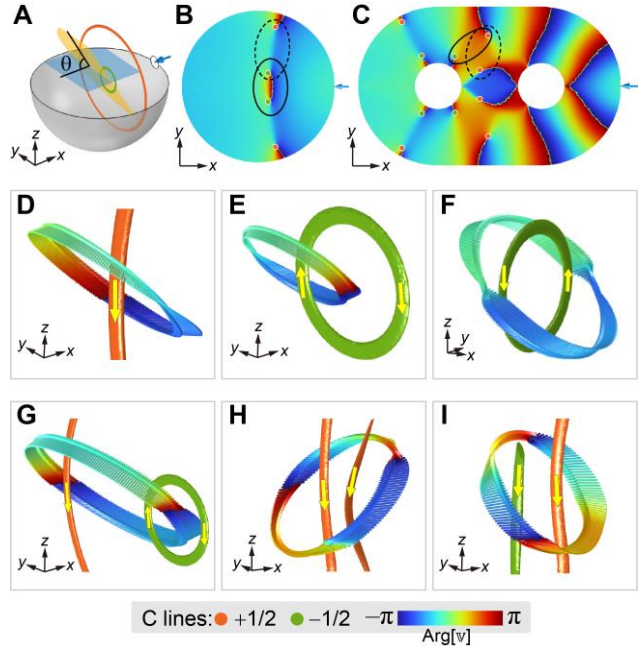

**Figure 5**. **Polarization Möbius strips associated with the C lines.** (**A**) the blue horizontal plane at z=0, and the yellow plane with a tilting angle $\theta$ on which the polarization Möbius strips are studied. (**B**), (**C**) Closed loops chosen in the spherical cavity and double-torus cavity for plotting the polarization major axes **A**. The background color depicts the phases Arg[$\mathbb{v}$]. (**D**), (**E**) Polarization major axes **A** on a loop enclosing the single orange C line and green C line in the spherical cavity, respectively. (**F**), (**G**) Polarization major axes **A** on the solid-line loop and dashed-line loop in (**B**), respectively. (**H**), (**I**) Polarization major axes **A** on the dashed-line loop and solid-line loop in (**C**), respectively. The color of the small arrows (i.e., the major axes **A**) indicates the phase Arg [$\mathbb{v}$]. The large yellow arrows on the C lines indicate the direction of the C lines.

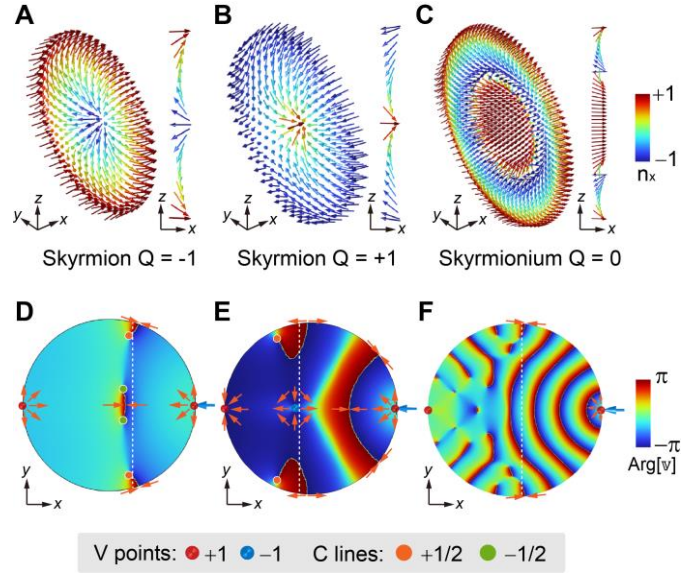

**Figure 6. Acoustic skyrmions and skyrmionium of velocity fields**. (**A**) Skyrmion with Q = −1. (**B**) Skyrmion with Q = +1. (**C**) Skyrmionium with Q = 0. The velocity vectors along the radial direction are depicted on the right side. The color of the arrows denotes the value of the *x*-component of the velocity vector. The frequency is 150 Hz for (**A**), 230 Hz for (**B**), and 600 Hz for (**C**). (**D**), (**E**), (**F**) Distribution of the phase Arg[$\mathbb{v}$] and polarization singularities on the *xoy*-plane of the spherical cavity in (**A**), (**B**), and (**C**), respectively. The white dashed lines mark the planes on which the skyrmions and skyrmionium locates. The orange arrows show the local velocity vectors.

# Sculpturing sound fields with real-space topology

Qing Tong[1] and Shubo Wang[1,*]

[1]Department of Physics, City University of Hong Kong, Tat Chee Avenue, Kowloon, Hong Kong, China.

*Corresponding author: Shubo Wang (shubwang@cityu.edu.hk)

## Supplementary Note 1: Polarization singularities in closed cavities and lossless cavities

In the main text, we focus on lossy acoustic cavities with a small opening (where the incident wave enters). The physics equally applies to closed cavities and lossless cavities with smooth surfaces.

Akin to the lossy open cavities, both C lines and V points can emerge inside lossy closed cavities under the excitation of a monopole source locating inside the cavities, as shown in Fig. S1. The cavities have the same material properties and geometric dimensions as those in Fig. 1 of the main text. The monopole source is denoted by a yellow dot. The distribution and configurations of the velocity polarization singularities are similar to those emerging in the open cavities in Fig. 1. We find that the total index of the velocity polarization singularities on the surface are $\sum_i I_i = 2, 0$, and -2, for spherical, torus and double-torus cavities, respectively. Therefore, the global topological properties of the velocity polarization field are also decided by the Euler characteristic of the cavities.

In contrast to the lossy open cavities which have both C lines and V points inside, only V points can emerge in the lossless open cavities due to the standing wave nature of sound inside the cavities. Figure S2 shows the polarization singularities emerging in the spherical, torus, and double-torus cavities, respectively. The geometric parameters and excitations of the cavities are the same as the cases in Fig. 1 of the main text. As seen, only V points emerge on the surfaces of the cavities. It is easy to verify that their total polarization index are $\sum_i I_i = 2, 0$, and -2, respectively, for the spherical, torus and double-torus cavities.

Thus, the global topological property of the polarization singularities in the lossless open cavities satisfy the PH theorem as in the lossy cases.

In addition, we verified the effect of geometric symmetry on the properties of the polarization singularities in the lossless open cavities. The results are shown in Fig. S3 for the lossless cavities with the same geometry as in Fig. 3(**A**) and 3(**B**) of the main text, which have Euler characteristics $\chi = 2$. As seen, in contrast to the lossy cavities, the two lossless cavities only generate V points with the total index $\sum_i I_i = 2 = \chi$. Thus, the global topological property of the V points is robust against continuous deformation of the cavities' geometry and is irrelevant to the symmetrical properties of the cavities.

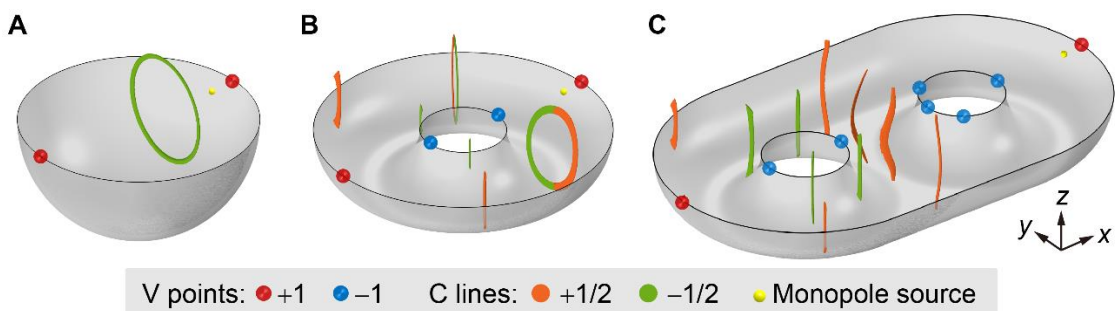

**Fig. S1** Polarization singularities generated in the lossy closed acoustic cavities with genus $g = 0$ for (**A**), $g = 1$ for (**B**), and $g = 2$ for (**C**) at $f = 150$ Hz. The yellow dots denote the positions of the monopole source.

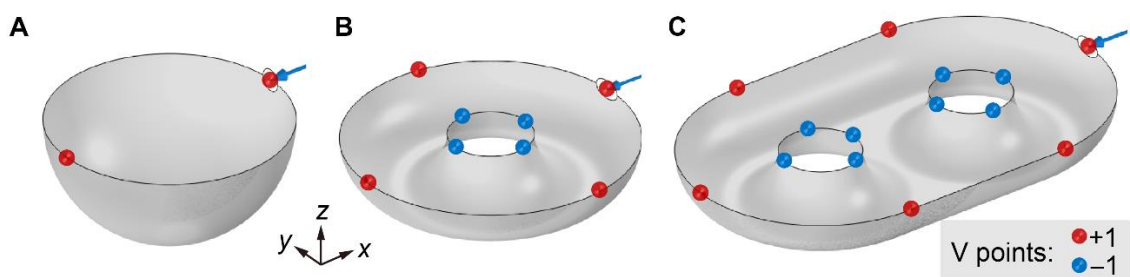

**Fig. S2** Polarization singularities generated in the lossless acoustic cavities with genus $g = 0$ for (**A**), $g = 1$ for (**B**), and $g = 2$ for (**C**) at $f = 100$ Hz. The sound speed is set to be 343 m/s. The red (blue) dots denote the V points with the polarization index $I_{pl} = +1$ ($I_{pl} = -1$). The blue arrow denotes the direction of incident waves at the opening.

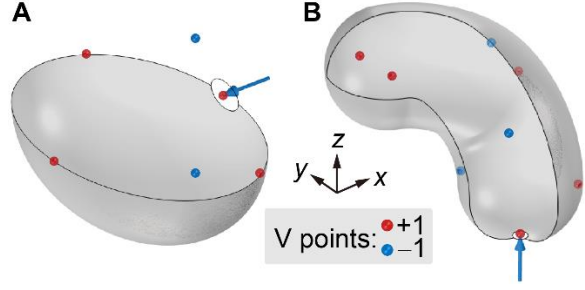

**Fig. S3** Polarization singularities generated in the lossless cavities (**A**) with a mirror symmetry and (**B**) without symmetry at $f = 150$ Hz. The V points on the surface are denoted by red (for polarization index $I_{pl} = +1$) and blue dots (for polarization index $I_{pl} = -1$).

## Supplementary Note 2: V point with anisotropic polarization index

Interesting phenomena can happen in the evolution of polarization singularities when the excitation frequency changes. As discussed in the main text regarding Fig. 4(**E**), the C ring with index $I_{pl} = -1/2$ converges to a V point locating inside the spherical cavity. This V point has an anisotropic polarization index: $I_{pl} = -1$ on any mirror plane (e.g., *xoy*-plane) and $I_{pl} = +1$ on the *yz*-plane. To understand this property, we show the polarization major axes **A** and the phase Arg [$v$] on the *xoy*-plane and *yz*-plane in Fig. S4. As shown in Fig. S4(**B**), the V point (denoted by the blue dot) clearly carries an index of $I_{pl} = -1$ on the *xoy*-plane. However, in the *yz*-plane shown in Fig. S4(**C**), the major axes **A** shows that the V point has index $I_{pl} = +1$. This anisotropic property of the polarization index of the V point is attributed to the symmetry of the system. The cylindrical symmetry of the spherical cavity leads to the polarization index $I_{pl} = +1$ on the *yz*-plane, transverse to the wave incident direction. On the *xoy*-plane, the merging of the two C points with the same polarization index $I_{pl} = -1/2$ but opposite phase index $I_{ph} = \pm 1$ (as shown by the phase Arg [$v$] and singularities in Fig. 4(**D**) of main text) necessarily gives rise to a V point with polarization index $I_{pl} = (-1/2) + (-1/2) = -1$ and phase index $I_{ph} = +1 - 1 = 0$, when increasing frequencies from 180 to 230 Hz.

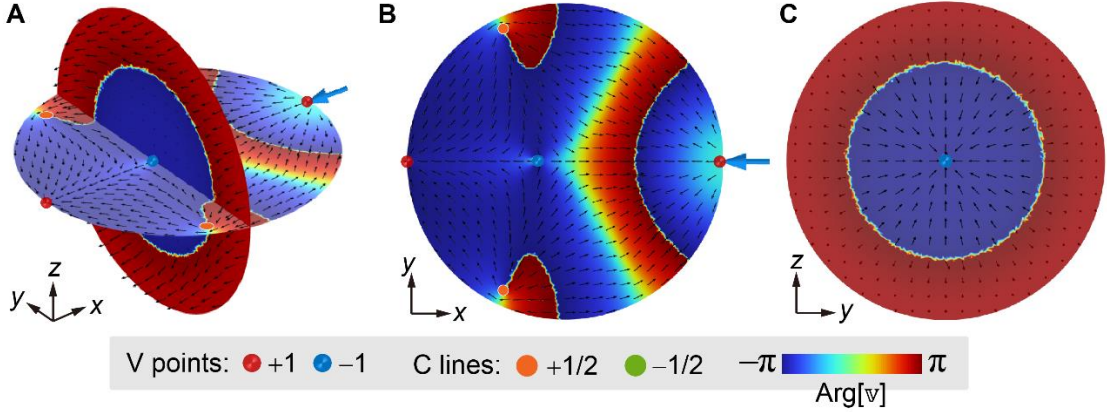

**Fig. S4** (**A**) The distributions of the polarization major axis **A** on two different cutting planes of the spherical cavity at $f = 230$ Hz (corresponding to the case in Fig. 4(**E**) of the main text). (**B**) and (**C**) correspond to the horizontal *xoy*-plane and the vertical *yz*-plane in (**A**), respectively.

## Supplementary Note 3: Time variation of the skyrmion number

In the main text, we show that the velocity vector can form intriguing skyrmion textures inside the cavity. These topological textures are time dependent and their skyrmion numbers generally change with time. Figure S5 presents the variation of the skyrmion number Q over a period of time for the skyrmion emerging in both the lossy and the lossless spherical cavities. We consider the skyrmion at $x = 0.267$ m in the lossy cavity at angular frequency $\omega_0 = 300\pi$, corresponding to the case in Fig. 6(**A**) of the main text. The skyrmion number Q is shown in Fig. S5(**A**) as a function of time, which exhibits a shape of step function with a periodic change of the value: $-1 \to 0 \to +1 \to 0 \to -1$. The small peaks of the skyrmion number near the transition points can be attributed to the variation of the phase Arg[$\mathbb{v}$] over time. In the bottom of Fig. S5(**A**), we show the skyrmion textures of the velocity field corresponding to different values of Q at different time.

The time dependence of the skyrmion number Q also happens in the lossless spherical cavity. We consider the skyrmion at $x = 0.20$ m in the cavity at angular frequency $\omega_0 = 500\pi$. The skyrmion number Q is shown in Fig. S5(**B**) as a function of time, which exhibits a periodic change of the value: $+1 \to -1 \to +1$. In contrast to the lossy case, the skyrmion number does not show small peaks at transition points due to the absence of C lines (and thus the lack of variation of phase Arg[v] over time). In the bottom of Fig. S5(**B**), we show the distribution of velocity vectors corresponding to different values of Q at different time.

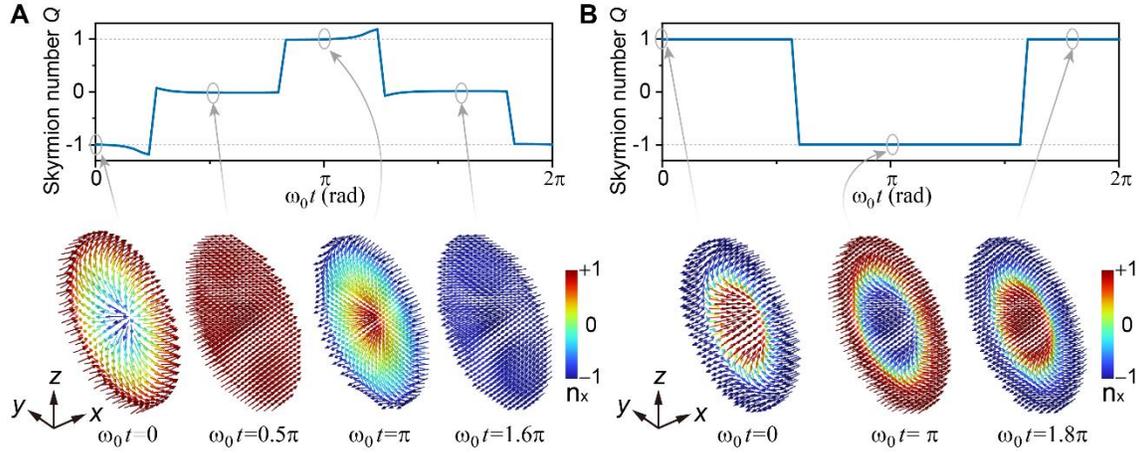

**Fig. S5** (**A**) The skyrmion number of acoustic velocity textures at $x = 0.267$ m and $f = 150$ Hz ($\omega_0 = 300\pi$) as a function of time for the lossy spherical cavity. Four velocity vector textures at different time are depicted in the bottom. (**B**) The skyrmion number of the acoustic velocity texture at $x = 0.2$ m and $f = 250$ Hz ($\omega_0 = 500\pi$) as a function of time for the lossless spherical cavity. Three velocity vector textures at different time are presented in the bottom.